\documentclass[aps,prx,twocolumn,amsmath,amssymb]{revtex4}

\usepackage{graphicx}
\usepackage{dcolumn}
\usepackage{bm}
\usepackage{epsfig}
\usepackage{amsmath}
\usepackage{amsfonts}

\newcommand{\ba}{\begin{eqnarray}}
\newcommand{\ea}{\end{eqnarray}}
\newcommand{\be}{\begin{equation}}
\newcommand{\ee}{\end{equation}}
\newcommand{\bea}{\begin{eqnarray}}
\newcommand{\eea}{\end{eqnarray}}

\usepackage{theorem}

\theorembodyfont{\rmfamily}

\theoremstyle{break}

\def\QED{~\rule[-1pt]{5pt}{5pt}\par\medskip}

\begin{document}


\title{Quantum parameter estimation with general dynamics}
\author{Haidong Yuan}
\email{hdyuan@mae.cuhk.edu.hk}
\affiliation{Department of Mechanical and Automation Engineering, The Chinese University of Hong Kong, Shatin, Hong Kong}

\author{Chi-Hang Fred Fung}
\email{chffung.app@gmail.com}
\affiliation{
Canada Research Centre, Huawei Technologies Canada, Ontario, Canada
}

\date{\today}

\begin{abstract}
One of the main quests in quantum metrology, and quantum parameter estimation in general, is to find out the highest achievable precision with given resources and design schemes that attain that precision. In this article we present a general framework for quantum parameter estimation which relates the ultimate precision limit directly to the geometrical properties of underlying dynamics. With this framework we present systematical methods for computing the ultimate precision limit and optimal probe states. We further demonstrate the power of the framework by deriving a sufficient condition on when ancillary systems are not useful for improving the precision limit.

\end{abstract}
\maketitle
\section{Introduction}
An important task in science and technology is to find the highest achievable precision in measuring and estimating parameters of interest with given resources, and design schemes to reach it. Quantum metrology, which exploits quantum mechanical effects to achieve high precision, has gained increasing attention in recent years\cite{Giovannetti2011, wineland,cavesprd,rosetta,VBRAU92-1,GIOV04,Fujiwara2008,Escher2011,Tsang2013,Rafal2012,durkin,Knysh2014,Jan2013,Rafal2014,Alipour2014,Chin2012,Tsang2011,Berry2013,Berry2015},
where a typical situation is to estimate the value of a continuous parameter $x$ encoded in some quantum state $\rho_x$ of the system. To estimate the value, one needs to first perform measurements on the system, which, in the general form, are described by Positive Operator Valued Measurements(POVM), $\{E_y\}$, which provides a distribution for the measurement results $p(y|x)=Tr(E_y\rho_x)$. According to the Cram\'{e}r-Rao bound in statistical theory\cite{HELS67,HOLE82,CRAM46,Rao}, the standard deviation for any unbiased estimator of $x$, based on the measurement results $y$, is bounded below by the Fisher information: $\delta \hat{x}\geq \frac{1}{\sqrt{I(x)}},$ where $\delta \hat{x}$ is the standard deviation of the estimation of $x$, and $I(x)$ is the Fisher information of the measurement results,
$I(x)=\sum_y p(y|x)(\frac{\partial lnp(y|x)}{\partial x})^2$\cite{Fisher}.
The Fisher information can be further optimized over all POVMs, which gives
\begin{equation}
\label{eq:J}
\delta\hat{x}\geq\frac{1}{\sqrt{\max_{E_y}I(x)}}=\frac{1}{\sqrt{J(\rho_x)}},
\end{equation}
 where the optimized value $J(\rho_x$) is called quantum Fisher information\cite{HELS67, HOLE82,BRAU94,BRAU96}.
If the above process is repeated $n$ times, then the standard deviation of the estimator is bounded by
$\delta \hat{x}\geq \frac{1}{\sqrt{nJ(\rho_x)}}.$

To achieve the best precision, we can further optimize the encoding procedures $x\rightarrow \rho_x$ so that $J(\rho_x)$ is maximized. Typically the encoding is achieved by preparing the probe in some initial state $\rho_0$, then let it evolve under a dynamics which contains the interested parameter, $\rho_0\xrightarrow{\phi_x} \rho_x$. Usually $\phi_x$ is determined by a given physical dynamics which is then fixed, while the initial state is up to our choice and can be optimized. A pivotal task in quantum metrology is to find out the optimal initial state $\rho_0$ and the corresponding maximum quantum Fisher information under any given evolution $\phi_x$.
When $\phi_x$ is unitary the GHZ-type of states are known to be optimal which leads to the Heisenberg limit. However when $\phi_x$ is noisy, such states are in general no longer optimal.
Finding the optimal probe states and the corresponding highest precision limit under general dynamics has been the main quest of the field.
Recently using the purification approach much progress has been made on developing systematical methods of calculating the highest precision limit\cite{Fujiwara2008,Escher2011,Tsang2013,Rafal2012,Jan2013,Rafal2014}, however how to actually achieve the highest precision limit is still largely unknown, as these methods do not provide ways to obtain the optimal probe states. Another restriction of these methods\cite{Fujiwara2008,Escher2011} is that they usually restrict to smooth representations of the Kraus operators, which is not intrinsic to the dynamics.

 In this article, we develop a general framework for quantum parameter estimation which relates the ultimate precision limit directly to the geometrical properties of underlying dynamics, this provides systematical methods for computing the ultimate precision limit and optimal probe states without additional assumptions. This framework also provides analytical formulas for the precision limit with arbitrary pure probe states which spares the needs of optimization over equivalent Kraus operators required in previous studies\cite{Fujiwara2008,Escher2011}. We further demonstrate the power of the framework by deriving sufficient conditions on when ancillary systems are not useful for improving the precision limit.


\section{Ultimate precision limit}
The precision limit of measuring $x$ from a set of quantum states $\rho_x$ is determined by the distinguishability between $\rho_x$ and its neighboring states $\rho_{x+dx}$\cite{BRAU94,Wootters1981}. This is best seen if we expand the Bures distance
 between the neighboring states $\rho_x$ and $\rho_{x+dx}$ up to the second order of $dx$\cite{BRAU94}:
\begin{equation}
\label{eq:BJ}
d^2_{Bures}(\rho_x,\rho_{x+dx})=\frac{1}{4}J(\rho_x)dx^2,
\end{equation}
where $d_{Bures}(\rho_1,\rho_2)=\sqrt{2-2F_B(\rho_1,\rho_2)}$, here $F_B(\rho_1,\rho_2)=Tr\sqrt{\rho_1^{\frac{1}{2}}\rho_2\rho_1^{\frac{1}{2}}}$ is the fidelity between two states.
Thus maximizing the quantum Fisher information is equivalent as maximizing the Bures distance, which is equivalent as minimizing the fidelity between $\rho_x$ and $\rho_{x+dx}$. If the evolution is given by $\phi_x$, $\rho_x=\phi_x(\rho)$ and $\rho_{x+dx}=\phi_{x+dx}(\rho)$, the problem is then equivalent to finding out $\min_{\rho}F_B[\phi_x(\rho),\phi_{x+dx}(\rho)]$ and the optimal $\rho$ that achieves the minimum. We now develop tools to solve this problem for both unitary and non-unitary dynamics.

Given two general evolution $\phi_1$ and $\phi_2$ of the same dimension, we define the Bures angle between them as $B(\phi_1,\phi_2)=\max_{\rho}\cos^{-1}[F_B(\phi_1(\rho),\phi_2(\rho))]$, this generalizes the Bures angle on quantum states\cite{Bures}. From the definition of the Bures distance it is easy to see $\max_{\rho} d^2_{Bures}[\phi_x(\rho),\phi_{x+dx}(\rho)]=2-2\cos B(\phi_x,\phi_{x+dx})$,
%
thus from Eq.(\ref{eq:BJ}) we have
\begin{eqnarray}
\label{eq:maxQFIphi1}
\aligned
\max_{\rho} J[\phi_x(\rho)]&=\lim_{dx\rightarrow 0}\frac{8[1-\cos B(\phi_x,\phi_{x+dx})]}{dx^2}.
\endaligned
\end{eqnarray}
The ultimate precision limit under the evolution $\phi_x$ is thus determined by the Bures angle between $\phi_x$ and the neighboring channels
\begin{equation}
\label{eq:Precision1}
\delta \hat{x}\geq\frac{1}{\lim_{dx\rightarrow 0} \frac{\sqrt{8[1-\cos B(\phi_x,\phi_{x+dx})]}}{\mid dx\mid}\sqrt{n}},
\end{equation}
where $n$ is the number of times that the procedure is repeated.
If $\phi_x$ is continuous with respect to $x$, then when $dx\rightarrow 0$, $B(\phi_x,\phi_{x+dx})\rightarrow B(\phi_x,\phi_{x})=0$, in this case
\begin{eqnarray}
\label{eq:maxQFIphi}
\aligned
\max_{\rho} J[\phi_x(\rho)]&=\lim_{dx\rightarrow 0}\frac{8[1-\cos B(\phi_x,\phi_{x+dx})]}{dx^2}\\
&=\lim_{dx\rightarrow 0}\frac{16\sin^2\frac{B(\phi_x,\phi_{x+dx})}{2}}{dx^2}\\
&=\lim_{dx\rightarrow 0}\frac{4B^2(\phi_x,\phi_{x+dx})}{dx^2},
\endaligned
\end{eqnarray}
the ultimate precision limit is then given by
\begin{equation}
\label{eq:Precision}
\delta \hat{x}\geq\frac{1}{\lim_{dx\rightarrow 0} 2\frac{B(\phi_x,\phi_{x+dx})}{\mid dx\mid}\sqrt{n}}.
\end{equation}

The problem is thus reduced to determine the Bures angle between quantum channels. We will first show how to compute the Bures angle between unitary channels, then generalize to noisy quantum channels.
\subsection{Ultimate precision limit for unitary channels}
Given two unitaries $U_1$ and $U_2$ of the same dimension, since $F_B(U_1\rho U^\dagger_1,U_2\rho U^\dagger_2)=F_B(\rho,U^\dagger_1U_2\rho U^\dagger_2U_1)$, we have $B(U_1,U_2)=B(I,U^\dagger_1U_2)$, i.e., the Bures angle between two unitaries can be reduced to the Bures angle between Identity and a unitary. For a $m\times m$ unitary matrix $U$, let $e^{-i\theta_j}$ be the eigenvalues of $U$, where $\theta_j\in(-\pi,\pi]$ for $1\leq j\leq m$, which we will call the eigen-angles of $U$. If $\theta_{\max}=\theta_1\geq \theta_2\geq \cdots \geq \theta_m=\theta_{\min}$ are arranged in decreasing order, then $B(I,U)=\frac{\theta_{\max}-\theta_{\min}}{2}$ when $\theta_{\max}-\theta_{\min}\leq \pi$\cite{ChildsPR00, Acin01, Duan2007,Chau2011,Fung2,Fung3}, specifically if $U=e^{-iHt}$, then $B(I,U)=\frac{(\lambda_{\max}-\lambda_{\min})t}{2}$ if $(\lambda_{\max}-\lambda_{\min})t\leq \pi$, where $\lambda_{\max(\min)}$ is the maximal (minimal) eigenvalue of $H$. This provides ways to compute Bures angles on unitary channels. For example, suppose the evolution takes the form $U(x)=(e^{-ixHt})^{\otimes N}$ (tensor product of $e^{-ixHt}$ for $N$ times, which means the same unitary evolution $e^{-ixHt}$ acts on all $N$ probes). Then
\begin{eqnarray}
\aligned
B[U(x),U(x+dx)]&=B[I,U^\dagger(x)U(x+dx)]\\
                &=B[I,(e^{-iHtdx})^{\otimes N}].\\
\endaligned
\end{eqnarray}
It is easy to see that the difference between the maximal eigen-angle and the minimal eigen-angle of $(e^{-iHtdx})^{\otimes N}$ is $\theta_{\max}-\theta_{\min}=N(\lambda_{\max}\mid dx\mid t-\lambda_{\min}\mid dx\mid t)$.
Thus $B(I,(e^{-iHtdx})^{\otimes N})=\frac{\theta_{\max}-\theta_{\min}}{2}=\frac{(N\lambda_{\max}\mid dx\mid-N\lambda_{\min}\mid dx\mid)t }{2},$ Eq.(\ref{eq:Precision}) then recovers the Heisenberg limit $$\delta \hat{x}\geq \frac{1}{\sqrt{n}(\lambda_{\max}-\lambda_{\min})t}\frac{1}{N}.$$
%

\subsection{Ultimate precision limit for noisy quantum channels}
For a general quantum channel which maps from a $m_1$- to $m_2$-dimensional Hilbert space, the evolution can be represented by a Kraus operation $K(\rho^S)=\sum_{j=1}^d F_j\rho^S F^\dagger_j$, here
the Kraus operators $F_j, 1\leq j\leq d$, are of the size $m_2\times m_1$, $\sum_{j=1}^d F^\dagger_jF_j=I_{m_1}$. The channel can be equivalently represented as
\begin{eqnarray}
\aligned
K(\rho^S)=Tr_E(U_{ES}(|0_E\rangle\langle0_E|\otimes \rho^S) U^\dagger_{ES}),\\
\endaligned
\end{eqnarray}
where $|0_E\rangle$ denotes some standard state of the environment, and $U_{ES}$ is a unitary operator acting on both system and environment, which we will call as the unitary extension of $K$.
A general $U_{ES}$ can be written as
\begin{align}
\label{eqn-U-general}
U_{ES}=(W_E \otimes I_{m_2})
\underbrace{
\begin{bmatrix}
F_1 & * & * & \cdots & *
\\
F_2 & * & * & \cdots & *
\\
\vdots & &\vdots & & \vdots
\\
F_{d} & * & * & \cdots & *\\
\textbf{0} & * & * & \cdots & *\\
\vdots & &\vdots & & \vdots\\
\textbf{0} & * & * & \cdots & *\\
\end{bmatrix}
}_{\displaystyle U},
\end{align}
here only the first $m_1$ columns of $U$ are fixed and $W_E \in U(p)$($p\times p$ unitaries) only acts on the environment and can be chosen arbitrarily, here $p\geq d$ as $p-d$ zero Kraus operators can be added.


Given a channel an ancillary system can be used to improve the precision limit, this can be described as the extended channel $$(K\otimes I_A) (\rho^{SA})=\sum_j (F_j\otimes I_A) \rho^{SA} (F_j\otimes I_A)^\dagger,$$ here $\rho^{SA}$ represents a state of the original and ancillary systems. Without loss of generality, the ancillary system can be assumed to have the same dimension as the original system.

Given two quantum channels $K_1$ and $K_2$ of the same dimension, let $U_{ES1}$ and $U_{ES2}$ as unitary extensions of $K_1$ and $K_2$ respectively, we have\cite{Yuan2015}
\begin{eqnarray}
\aligned
B(K_1\otimes I_A,K_2\otimes I_A)&=\min_{U_{ES1},U_{ES2}}B(U_{ES1},U_{ES2})\\
&=\min_{U_{ES1}}B(U_{ES1},U_{ES2})\\
&=\min_{U_{ES2}}B(U_{ES1},U_{ES2}).
\endaligned
\end{eqnarray}
This extends Uhlmann's purication theorem on mixed states\cite{Uhlmann1976} to noisy quantum channels.
Furthermore we show in the appendix that $B(K_1\otimes I_A,K_2\otimes I_A)$ can be explicitly computed from the Kraus operators of $K_1$ and $K_2$\cite{supplement}: if $K_1(\rho^S)=\sum_{j=1}^d F_{1j}\rho^S F^\dagger_{1j}$, $K_2(\rho^S)=\sum_{j=1}^d F_{2j}\rho^S F^\dagger_{2j}$, then
$\cos B(K_1\otimes I_A, K_2\otimes I_A) =\max_{\|W\|\leq 1}\frac{1}{2}\lambda_{\min}(K_W+K^\dagger_W),$  here $\lambda_{\min}(K_W+K^\dagger_W)$ denotes the minimum eigenvalue of $K_W+K^\dagger_W$ where $K_W=\sum_{ij}w_{ij}F^\dagger_{1i}F_{2j}$, with $w_{ij}$ as the $ij$-th entry of a $d\times d$ matrix $W$ which satisfies $\|W\|\leq 1$($\|\cdot\|$ denotes the operator norm which equals to the maximum singular value). If we substitute $K_1=K_x$ and $K_2=K_{x+dx}$, where $K_x(\rho^S)=\sum_{j=1}^d F_j(x)\rho^S F^\dagger_j(x)$ and $K_{x+dx}(\rho^S)=\sum_{j=1}^d F_j(x+dx)\rho^SF^\dagger_j(x+dx)$ with $x$ being the interested parameter, then
\begin{eqnarray}
\aligned
\label{eq:suppTE}
&\cos B(K_x\otimes I_A, K_{x+dx}\otimes I_A)\\
=& \max_{\|W\|\leq 1 }\frac{1}{2}\lambda_{\min}(K_W+K^\dagger_W),
 \endaligned
 \end{eqnarray}
where $K_W=\sum_{ij}w_{ij}F^\dagger_i(x)F_j(x+dx)$.
By substituting $\phi_x=K_x\otimes I_A$ and $\phi_{x+dx}=K_{x+dx}\otimes I_A$ in Eq.(\ref{eq:maxQFIphi1}), we then get the maximal quantum Fisher information for the extended channel $K_x\otimes I_A$,
\begin{eqnarray}
\aligned
\label{eq:maxQFIopen}
\max J=\lim_{dx\rightarrow 0}\frac{8[1-\max_{\|W\|\leq 1 }\frac{1}{2}\lambda_{\min}(K_W+K^\dagger_W)]}{dx^2}.
\endaligned
\end{eqnarray}

In previous studies the operator $W_E$ in Eq.(\ref{eqn-U-general}), which can be arbitrary chosen, was assumed to depend on $x$ smoothly\cite{Fujiwara2008,Escher2011}. As a result, the $W$ in Eq.(\ref{eq:maxQFIopen}) was restricted to unitary operators that depends smoothly on $x$ as explained in detail in the appendix \ref{sec:con}. This restriction was introduced out of computational convenience in previous studies, which is not intrinsic to the dynamics.
 The formula here does not have such assumption and can be applied more broadly, for example it can be applied to the discrimination of quantum channels which is discrete in nature\cite{Yuan2015}. Also since any $W$ that is not optimal gives a lower bound on the precision limit, the formula here also provides more room for obtaining useful lower bounds.

The maximization in Eq.(\ref{eq:maxQFIopen}) can be further formulated as a semi-definite programming and solved efficiently:  $\max_{\|W\|\leq 1}\frac{1}{2}\lambda_{\min}(K_W+K^\dagger_W)=$
\begin{eqnarray}
\label{eq:sdp}
\aligned
&maximize \qquad \frac{1}{2}t \\
s.t.\qquad &\left(\begin{array}{cc}
      I & W^\dagger  \\
      W & I \\
          \end{array}\right)\succeq 0,\\
      &    K_W+K^\dagger_W-tI \succeq 0.
          \endaligned
          \end{eqnarray}

 Another advantage of this formulation is that the dual form of this semi-definite programming provides a systematical way for obtaining the optimal probe states, which we will show in the next setion. 


\section{Optimal probe states}
Developing systematical methods to obtain the optimal probe states are essential for achieving the precision limit. So far there are only a few cases for which optimal probe states are known, mostly for phase estimations\cite{durkin,Berry2000,Rafal2009, Nair2011,Knysh2014, Frowis2014}. 
A systematical way of obtaining optimal probe states for general quantum dynamics is highly desired as it will pave the way for achieving the ultimate precision limit. We now show how to obtain the optimal probe states that achieve the ultimate precision limit for extended channels.

We first provide an analytical formula for calculating quantum Fisher information with any given pure input states, for both unextended and extended channels, then use it to obtain optimal probe states for extended channels.

In the appendix we showed that for both unextended and extended channels with pure probe states we have\cite{supplement}
\begin{equation}
\label{eq:fidelityS}
F_B[K_x(\rho^{S}),K_{x+dx}(\rho^{S})]=\|M(\rho^S)\|_1,
\end{equation}
\begin{equation}
\label{eq:fidelitySA}
F_B[(K_x\otimes I_A)(\rho^{SA}),(K_{x+dx}\otimes I_A)(\rho^{SA})]=\|M(\rho^S)\|_1,
\end{equation}
here $M(\rho^S)$ is a $d\times d$ matrix with its $ij$-entry equals to $Tr[\rho^S F^\dagger_i(x)F_j(x+dx)]$, and $\|\cdot\|_1$ represents the trace norm which equals to the summation of singular values. For the unextended channel this formula works for the pure probe state $\rho^S=|\psi_S\rangle\langle \psi_S|$, while for the extended channel although $\rho^{SA}=|\psi_{SA}\rangle\langle \psi_{SA}|$ is required to be a pure state, $\rho^S=Tr_A(\rho^{SA})$ can be any mixed state, which characterizes the advantage provided by ancillary systems.

The above two formulas provide a straightforward way calculating the quantum Fisher information with any pure probe states,
\begin{eqnarray}
\label{eq:QFIM}
\aligned
J[K_x(\rho^{S})]&=\lim_{dx\rightarrow 0}\frac{8(1-\|M(\rho^S)\|_1)}{dx^2},\\
J[(K_x\otimes I_A)(\rho^{SA})]&=\lim_{dx\rightarrow 0}\frac{8(1-\|M(\rho^S)\|_1)}{dx^2}.
\endaligned
\end{eqnarray}
In contrast to previous studies\cite{Fujiwara2008,Escher2011,Escher2012}, optimization over equivalent representations of Kraus operator is not needed in this formulation. In fact $\|M(\rho^S)\|_1$ does not depend on any particular representation of the Kraus operators: if we use a different representation of Kraus operators for $K_x$ and $K_{x+dx}$, for example $\tilde{F}_i(x)=\sum_r u_{ir}F_r(x)$ and  $\tilde{F}_j(x+dx)=\sum_s v_{js}F_s(x+dx)$, where $u_{ir}$ and $v_{js}$ are entries of some unitary matrices $U$ and $V$ respectively, then $Tr[\rho^S \tilde{F}^\dagger_i(x)\tilde{F}_j(x+dx)]=\sum_{r,s} u^*_{ir}v_{js}Tr[\rho^S F^\dagger_r(x)F_s(x+dx)]$, thus $\tilde{M}(\rho^S)=\bar{U} M(\rho^S)V^T$ which has the same trace norm $\|\tilde{M}(\rho^S)\|_1=\|M(\rho^S)\|_1$. 

%

The optimal probe states for the extended channel can then be obtained by minimizing over input states at both sides of Eq.(\ref{eq:fidelitySA}),
\begin{eqnarray}
\label{eq:optimalstate}
\aligned
&\min_{\rho^{SA}}F_B[(K_x\otimes I_A)(\rho^{SA}),(K_{x+dx}\otimes I_A)(\rho^{SA})]\\
=&\min_{\rho^S}\|M(\rho^S)\|_1.
\endaligned
\end{eqnarray}

This can be computed by a semi-definite programming formulation for the trace norm\cite{Fuel} as
$\min_{\rho^S}\|M(\rho^S)\|_1=$
\begin{eqnarray}
\label{eq:SDPrho}
\aligned
minimize \qquad &\frac{1}{2}Tr(P)+\frac{1}{2}Tr(Q) \\
s.t.\qquad &\left(\begin{array}{cc}
      P & M^\dagger(\rho^S)  \\
      M(\rho^S) & Q \\
          \end{array}\right)\succeq 0,\\
       & \rho^S\succeq 0,
        Tr(\rho^S)=1,
          \endaligned
          \end{eqnarray}
where $P, Q$ are Hermitian matrices. One can verify that this is exactly the dual form of the semi-definite programming used in Eq.(\ref{eq:sdp}). From the output $\rho^S$, we can easily obtain the optimal probe state $\rho^{SA}$, which is any purification with the reduced states equals to the optimal $\rho^S$. This gives a systematical way to obtain the optimal probe states for the extended channel which we demonstrate through some examples.

Consider phase estimation with spontaneous emission, $K_x(\rho_0)=F_1(x)\rho_0F^\dagger_1(x)+F_2(x)\rho_0F_2^\dagger(x)$, where $F_1(x)=\left(\begin{array}{cc}
      1 & 0  \\
      0 & \sqrt{\eta} \\
          \end{array}\right)U(x)$,$F_2(x)=\left(\begin{array}{cc}
      0 & \sqrt{1-\eta}  \\
      0 & 0 \\
          \end{array}\right)U(x)$,
$U(x)=\exp(-i\frac{\sigma_3}{2}x)$. Suppose a pure input state $\rho^{SA}$ is prepared for the extended channel $K_x\otimes I_A$, then $\rho^S=Tr_A(\rho^{SA})$,
$M(\rho^S)=\left(\begin{array}{cc}
      Tr[\rho^SF_1^\dagger(x)F_1(x+dx)] & Tr[\rho^SF_1^\dagger(x)F_2(x+dx)]  \\
      Tr[\rho^SF_2^\dagger(x)F_1(x+dx)] & Tr[\rho^SF_2^\dagger(x)F_2(x+dx)] \\
          \end{array}\right),$
in this case the problem can be solved analytically, in the appendix we showed that the optimal $\rho^S$ is given by $\rho^S=\left(\begin{array}{cc}
      \frac{\sqrt{\eta}}{1+\sqrt{\eta}} & 0  \\
      0 & \frac{1}{1+\sqrt{\eta}} \\
          \end{array}\right)$ and the corresponding maximal quantum Fisher information is $\max J=\frac{4\eta}{(1+\sqrt{\eta})^2}$\cite{supplement}. Since the optimal $\rho^S$ is mixed, an ancillary systems is necessary. The optimal input state in this case is any pure state $\rho^{SA}$ with the reduced state equal to $\rho^S$, the simplest choice of the optimal input state in this case is $\sqrt{\frac{\sqrt{\eta}}{1+\sqrt{\eta}}}|00\rangle+\sqrt{\frac{1}{1+\sqrt{\eta}}}|11\rangle$, which is not a maximally entangled state as previously suspected\cite{Jan2013,Jan2014}.
We can also use the method to find the maximal quantum Fisher information without using ancilla by imposing the condition that $\rho^S$ be pure. In that case the maximal quantum Fisher information turns out to be $\eta$ and the optimal input state has the form $(|0\rangle+\exp(i \theta)|1\rangle)/\sqrt{2}$ for some $\theta \in \mathbb{R}$\cite{supplement}.

\begin{figure}
\centering
\begin{minipage}{.5\textwidth}
 \centering
  \includegraphics[scale=.4]{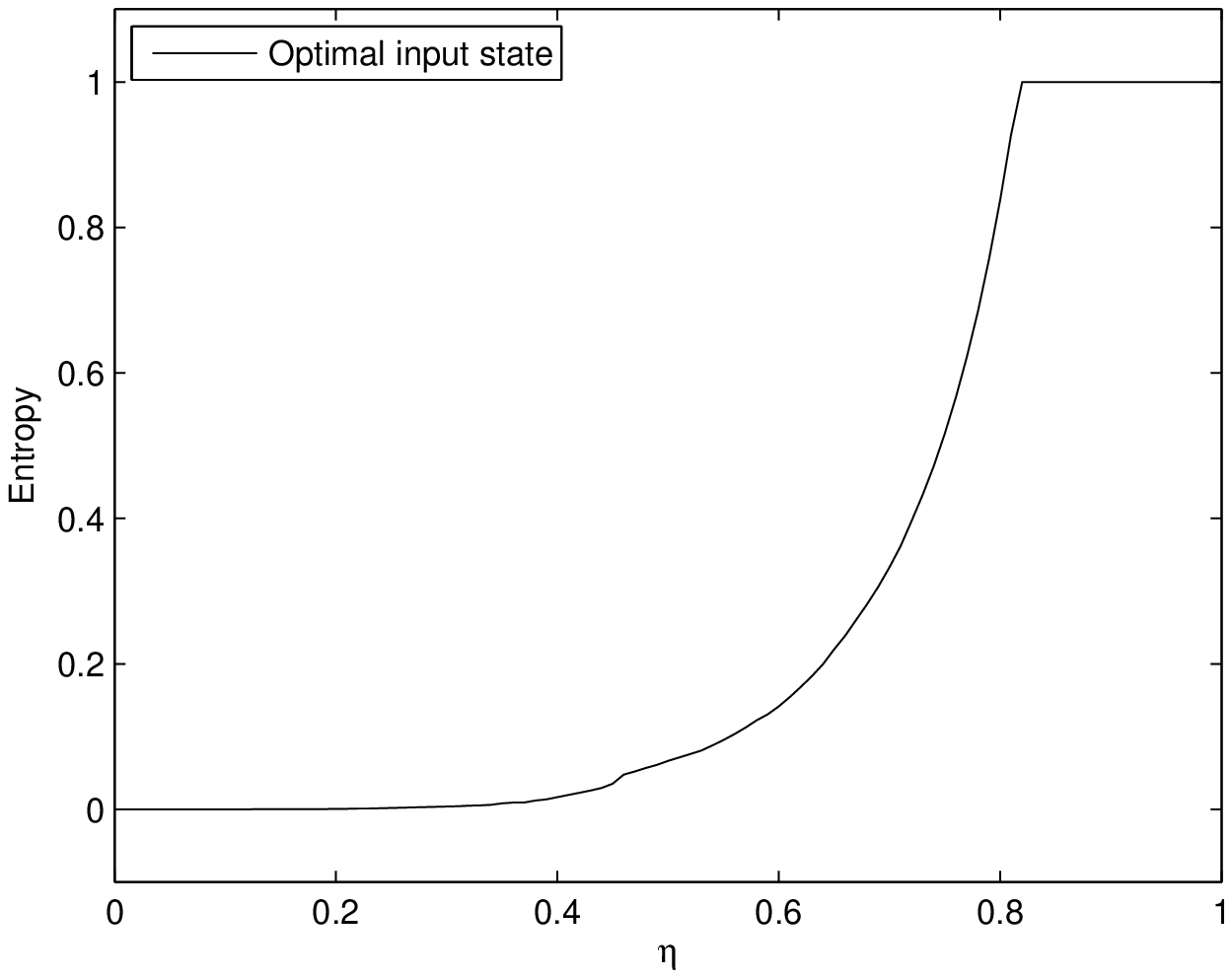}
  \caption{Entropy of reduced single qubit state, which is used to quantify the entanglement of the optimal state, at different $\eta$.}
  \label{fig:entropy}
\end{minipage}%
\qquad
\begin{minipage}{.5\textwidth}
  \centering
  \includegraphics[scale=.4]{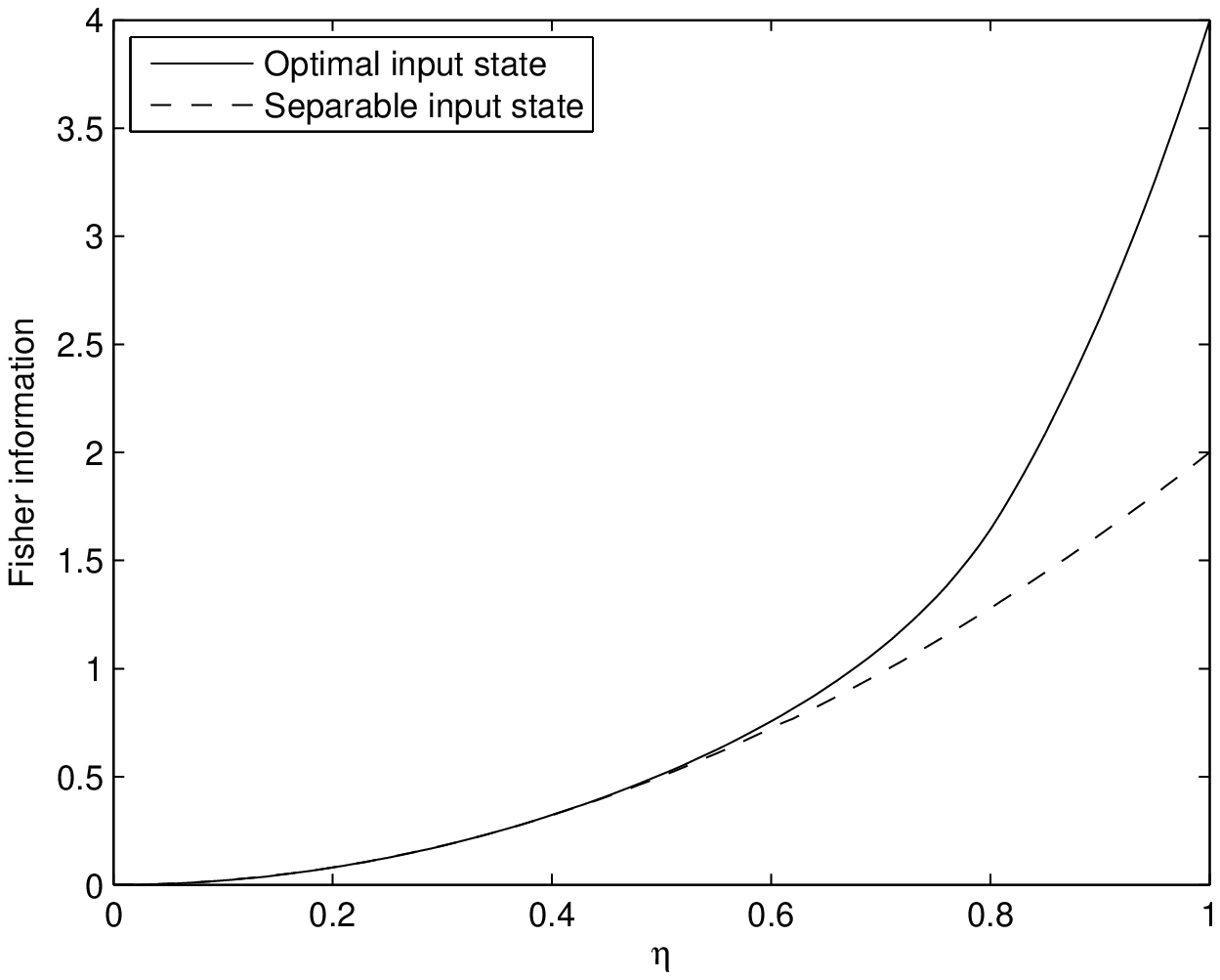}
  \caption{Quantum Fisher information with the optimal input state and separable input state $|++\rangle$ for 2 qubits with independent dephasing noises.}
  \label{fisher}
\end{minipage}
\end{figure}

\begin{figure}
  \centering
  \includegraphics[scale=.4]{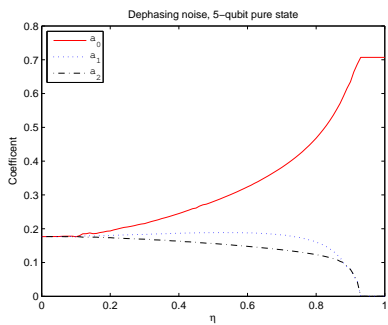}
  \caption{Optimal probe state for 5 qubits under independent dephasing noises. The optimal state has the form $|\psi\rangle=a_0|\psi_0\rangle+a_1|\psi_1\rangle+a_2|\psi_2\rangle$, where $|\psi_i\rangle$ denotes the summation of all basis states with $i$ zeros or $i$ ones, for example $|\psi_0\rangle=|00000\rangle+|11111\rangle$. }
  \label{fig:dep5}
\end{figure}

\begin{figure}
  \centering
  \includegraphics[scale=.4]{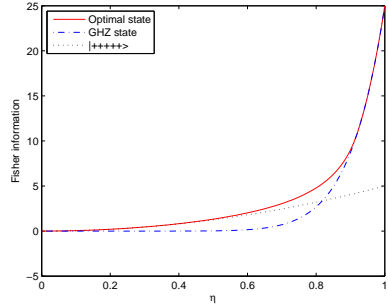}
  \caption{Quantum Fisher information for optimal probe states, GHZ state and separable state for 5 qubits under independent dephasing noises. }
  \label{fig:Dep5Fisher}
\end{figure}

\begin{figure}
  \centering
  \includegraphics[scale=.4]{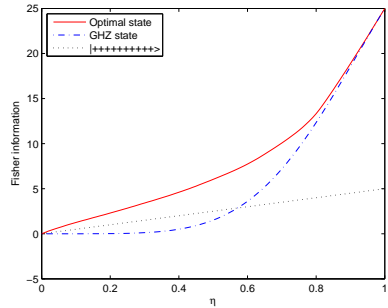}
  \caption{Quantum Fisher information for optimal probe states, GHZ state and separable state for 10 qubits with 5 of them under independent spontaneous emission and 5 of them as ancillary system.. }
  \label{fig:Sop10Fisher}
\end{figure}

For high dimensional systems, we use the CVX package in Matlab\cite{CVX} to implement the semi-definite programming of (\ref{eq:SDPrho}) and obtain the optimal input states. For example, consider two qubits with independent dephasing noises, which can be represented by a Kraus operation with 4-Kraus operators: $F_1(x)\otimes F_1(x), F_1(x)\otimes F_2(x), F_2(x)\otimes F_1(x), F_2(x)\otimes F_1(x)$ with $F_1(x)=\sqrt{\frac{1+\eta}{2}}U(x)$, $F_2(x)=\sqrt{\frac{1-\eta}{2}}\sigma_3U(x)$, here $U(x)=\exp(-i\frac{\sigma_3}{2}x).$ It turns out that $\min_{\rho^S}\|M(\rho^S)\|$ can always be attained with a pure $\rho^S$, ancillary systems are thus not necessary. In Fig.~\ref{fig:entropy} we plotted the entanglement of optimal states, which is quantified by the entropy of the reduced single qubit state $\rho^S$, at different $\eta$. It can be seen that there exists a threshold for $\eta$: when $\eta$ exceeds the threshold, the optimal state is the GHZ state, which is maximally entangled; when $\eta$ is below the threshold, GHZ state ceases to be optimal, with the decreasing of $\eta$, the optimal state gradually changes from the maximally entangled state to separable state.  Fig.~\ref{fisher} shows the quantum Fisher information with the optimal state and the separable input state $|++\rangle$, where $|+\rangle=\frac{|0\rangle+|1\rangle}{\sqrt{2}}$. It can be seen that the gain of entanglement is only obvious in the region of high $\eta$, i.e., low noises. Similar behaviour is found for more qubits, i.e., there exists a threshold for $\eta$, above the threshold the optimal state is the GHZ state and with the decreasing of $\eta$, the optimal state gradually changes from GHZ state to separable state, and this threshold increases with the number of qubits. 
In Fig.\ref{fig:dep5} the optimal state for 5 qubits with independent dephasing noises is shown, and in Fig.\ref{fig:Dep5Fisher} the quantum Fisher information for the optimal state, GHZ state and the separable state are plotted. We also calculated the optimal state for 10 qubits with 5 of them under independent spontaneous emission and 5 of them as ancillary qubits, in Fig.\ref{fig:Sop10Fisher} we plotted the Fisher information for the optimal state, GHZ state and separable state. 

\section{When ancillary systems are not helpful}
The formulas developed here not only provides systematical methods to compute the ultimate precision limit and optimal probe states, but also have wide implications, which we will demonstrate by deriving a sufficient condition on when ancillary systems are not useful for improving the precision limit. 

We have seen that in the spontaneous emission case ancillary system helps improving the precision limit while in some other cases---known examples include unitary, classical, phase estimation with dephasing and lossy channels\cite{Jan2013,Jan2014}, ancillary systems do not help. For a general channel it is usually difficult to tell whether ancillary systems can help improving the precision limit or not. Previously this problem was usually studied case by case by comparing the maximum quantum Fisher information for the unextended and extended channels. The formulas developed here provide a more direct way: from Eq.(\ref{eq:QFIM}) it is obvious that ancillary systems do not help improving the precision limit if and only if $\min_{\rho^S}\|M(\rho^S)\|_1$ can be reached at a pure state $\rho^S$, which can be checked by using the semi-definite programming to obtain optimal $\rho^S$ and see if $(\rho^S)^2=\rho^S$. A more easily verifiable sufficient condition is as following: given a channel $K_x(\rho)=\sum_{i=1}^d F_i(x)\rho F^\dagger_i(x)$, if $F_i^\dagger(x)F_j(x+dx)$, $1\leq i,j\leq d$ can be simultaneously diagonalized, then ancillary systems do not help improving the precision limit. As if there exist a basis such that $F_i^\dagger(x)F_j(x+dx)$ are all diagonal, then in that basis only the diagonal entries of $\rho^S$ enter into $M(\rho^S)$(as the entries of $M(\rho^S)$ are of the form $Tr[\rho^S F_i^\dagger(x)F_j(x+dx)]$ which only depends on the diagonal terms of $\rho^S$ if $F_i^\dagger(x)F_j(x+dx)$ is diagonal), other entries can be chosen freely. This means that $\min_{\rho^S}\|M(\rho^S)\|_1$ can be achieved by all $\rho^S$ with the optimal diagonal entries, $\rho^S_{ii}=a_i$, this always includes a pure state $|\psi_S\rangle=\sum_i \sqrt{a_i}|i\rangle$, hence ancillary system is not necessary to achieve $\min_{\rho^S}\|M(\rho^S)\|_1$ for such channels. This sufficient condition is satisfied by unitary, classical and phase estimation with dephasing channel, and many other channels that have not been categorized before, for example phase estimation with noises along the $X$ and $Y$ directions satisfies this condition, as it can be represented with two following Kraus operators $F_1(x)=\sqrt{\frac{1+\eta}{2}}\sigma_1\exp(-i\frac{\sigma_3}{2}x)$ and $F_2(x)=\sqrt{\frac{1-\eta}{2}}\sigma_2\exp(-i\frac{\sigma_3}{2}x)$, one can easily check that they satisfy the condition thus ancillary system does not help improving the ultimate precision limit for this channel.

%
\section{Conclusion}
In conclusion we presented a general framework for quantum metrology which provides systematical ways of obtaining the ultimate precision limit and optimal input states. This framework relates the ultimate precision limit directly to the underlying dynamics, which opens the possibility of utilizing quantum control methods to alter the underlying dynamics for better precision limit\cite{YuanTime}. The tools developed here, such as the generalized Bures angle on quantum channels that can be efficiently computed using semi-definite programming, are expected to find wide applications in various fields of quantum information science.
\appendix
\section{Appendix}
In the appendix we give detailed derivation on the Bures angle between extended channels, the analytical formula to compute quantum Fisher information with arbitrary pure input state and two examples with dephasing noise and spontaneous emission respectively.
\subsection{Bures angle between extended channels}

In this section we show that for any two given channels which maps from $m_1$- to $m_2$-dimensional Hilbert space, $K_1(\rho^S)=\sum_{i=1}^d F_{1i}\rho^S F^\dagger_{1i}$ and $K_2(\rho^S)=\sum_{i=1}^d F_{2i}\rho^S F^\dagger_{2i}$, the Bures angle between the extended channels $K_1\otimes I_A$ and $K_2\otimes I_A$ can be computed from the Kraus operators as following
$$\cos B(K_1\otimes I_A,K_2\otimes I_A)=\max_{\| W\|\leq 1}\frac{1}{2}\lambda_{\min}(K_W+K_W^\dagger),$$
 i.e.,
 \begin{eqnarray}
\aligned
 &\min_{\rho^{SA}} F_B[(K_1\otimes I_A)(\rho^{SA}), (K_2\otimes I_A)(\rho^{SA})]\\
 =&\max_{\| W\|\leq 1}\frac{1}{2}\lambda_{\min}(K_W+K_W^\dagger),
  \endaligned
 \end{eqnarray}
 here $K_W=\sum_{ij}w_{ij}F_{1i}^\dagger F_{2j}$, $w_{ij}$ is the $ij$-th entry of $d\times d$ matrix $W$ and the optimization is over all $d\times d$ matrices $W$ with operator norm $\|W\|\leq 1$. During the proof the analytical formulas that compute quantum Fisher information for arbitrary pure input states will be derived.

As the minimum fidelity can always be achieved with pure states, we can assume $\rho^{SA}=|\psi_{SA}\rangle\langle \psi_{SA}|$. Denote $U_{ES1}$, $U_{ES2}$ as the unitary extension of $K_1$ and $K_2$ respectively, i.e.,
\begin{eqnarray}
\aligned
K_1(\rho^S)=Tr_E(U_{ES1}(|0_E\rangle\langle0_E|\otimes \rho^S )U^\dagger_{ES1}),\\
K_2(\rho^S)=Tr_E(U_{ES2}(|0_E\rangle\langle0_E|\otimes \rho^S) U^\dagger_{ES2}),\\
\endaligned
\end{eqnarray}
where $|0_E\rangle$ denotes some standard state of the environment, $U_{ES1}$ and $U_{ES2}$ are unitary operators acting on both system and environment.
The general form of $U_{ES1}$  can be written as a $pm_2$-dimensional unitary, here $p\geq d$, $p-d$ zero Kraus operators can be added,
\begin{align}
\label{eqn-U-general-form}
U_{ES1}=(W_1 \otimes I_{m_2})
\underbrace{
\begin{bmatrix}
F_{11} & * & * & \cdots & *
\\
F_{12} & * & * & \cdots & *
\\
\vdots & &\vdots & & \vdots
\\
F_{1d} & * & * & \cdots & *\\
\textbf{0} & * & * & \cdots & *\\
\vdots & &\vdots & & \vdots\\
\textbf{0} & * & * & \cdots & *\\
\end{bmatrix}
}_{\displaystyle U_1}
\end{align}
where $W_1 \in U(p)$ and only the first $m_1$ columns of $U_1$ are fixed. Similarly
\begin{align}
\label{eqn-U-general-form2}
U_{ES2}=(W_2 \otimes I_{m_2})
\underbrace{
\begin{bmatrix}
F_{21} & * & * & \cdots & *
\\
F_{22} & * & * & \cdots & *
\\
\vdots & &\vdots & & \vdots
\\
F_{2d} & * & * & \cdots & *\\
\textbf{0} & * & * & \cdots & *\\
\vdots & &\vdots & & \vdots\\
\textbf{0} & * & * & \cdots & *\\
\end{bmatrix}
}_{\displaystyle U_2}
\end{align}
where $W_2 \in U(p)$ and only the first $m_1$ columns of $U_2$ are fixed. Note that $W_1$ and $W_2$ only act on the environment and can be chosen arbitrarily.
\begin{widetext}
\begin{eqnarray}
\aligned
U_{ES1}^\dagger U_{ES2}&=\begin{bmatrix}
F^\dagger_{11} & F^\dagger_{12} & * & \cdots & F^\dagger_{1d} & \textbf{0} & * &\cdots &\textbf{0}
\\
* & * & * & \cdots & * & * & * & \cdots & *
\\
\vdots & & & & \vdots  &  & &\vdots
\\
* & * & * & \cdots & *  & * & * & \cdots & *
\end{bmatrix}(W_1^\dagger \otimes I_{m_2})(W_2 \otimes I_{m_2})\begin{bmatrix}
F_{21} & * & * & \cdots & *
\\
F_{22} & * & * & \cdots & *
\\
\vdots & & & & \vdots
\\
F_{2d} & * & * & \cdots & *\\
\textbf{0} & * & * & \cdots & *\\
\vdots & &\vdots & & \vdots\\
\textbf{0} & * & * & \cdots & *\\
\end{bmatrix}\\
&=\begin{bmatrix}
F^\dagger_{11} & F^\dagger_{12} & * & \cdots & F^\dagger_{1d} & \textbf{0} & * &\cdots &\textbf{0}
\\
* & * & * & \cdots & * & * & * & \cdots & *
\\
\vdots & & & & \vdots &  &  & &\vdots
\\
* & * & * & \cdots & * & * & * & \cdots & *
\end{bmatrix}(W_p \otimes I_{m_2})\begin{bmatrix}
F_{21} & * & * & \cdots & *
\\
F_{22} & * & * & \cdots & *
\\
\vdots & & & & \vdots
\\
F_{2d} & * & * & \cdots & *\\
\textbf{0} & * & * & \cdots & *\\
\vdots & &\vdots & & \vdots\\
\textbf{0} & * & * & \cdots & *\\
\end{bmatrix}\\
&=\begin{bmatrix}
K_{W_d} & * & * & \cdots & *
\\
* & * & * & \cdots & *
\\
\vdots & & & & \vdots
\\
* & * & * & \cdots & *
\end{bmatrix},
\endaligned
\end{eqnarray}
\end{widetext}
where $W_p=W_1^\dagger W_2\in U(p)$, $W_d$ is the first $d\times d$ submatrix of $W_p$, i.e.,
\begin{eqnarray}
\label{eq:norm}
W_p=\begin{bmatrix}
W_d & *\\
* & *\\
\end{bmatrix}
\end{eqnarray}
and $K_{W_d}=\sum_{i=1}^d\sum_{j=1}^dw_{ij}F_{1i}^\dagger F_{2j}$ with $w_{ij}$ as the $ij$-th entry of $W_d$. Then
\begin{widetext}
\begin{eqnarray}
\nonumber
\aligned
&\min_{|\psi_{SA}\rangle} F_B[(K_1\otimes I_A)(|\psi_{SA}\rangle\langle \psi_{SA}|), (K_2\otimes I_A)(|\psi_{SA}\rangle\langle \psi_{SA}|)]\\
=&\min_{|\psi_{SA}\rangle} F_B[Tr_E(U_{ES1}\otimes I_A(|0_E\rangle\langle0_E|\otimes |\psi_{SA}\rangle\langle \psi_{SA}|) U^\dagger_{ES1}\otimes I_A), Tr_E(U_{ES2}\otimes I_A(|0_E\rangle\langle0_E|\otimes |\psi_{SA}\rangle\langle \psi_{SA}|) U^\dagger_{ES2}\otimes I_A)]\\
=&\min_{|\psi_{SA}\rangle}\max_{U_{ES2}}F_B(U_{ES1}\otimes I_A(|0_E\rangle\langle0_E|\otimes |\psi_{SA}\rangle\langle \psi_{SA}|) U^\dagger_{ES1}\otimes I_A,U_{ES2}\otimes I_A(|0_E\rangle\langle0_E|\otimes |\psi_{SA}\rangle\langle \psi_{SA}|) U^\dagger_{ES2}\otimes I_A),
\endaligned
\end{eqnarray}
\end{widetext}
the second equality used Uhlmann's theorem which states that
\begin{eqnarray}
F_B(\rho_1,\rho_2)=\max_{|\psi_2\rangle}F_B(|\psi_1\rangle \langle\psi_1|, |\psi_2\rangle \langle\psi_2|),
\end{eqnarray}
where $|\psi_1\rangle$ is an arbitrary purification of $\rho_1$ and the maximization runs over all purifications $|\psi_2\rangle$ of $\rho_2$\cite{Uhlmann1976}. Thus
\begin{widetext}
\begin{eqnarray}
\label{eq:prove}
\aligned
&\min_{|\psi_{SA}\rangle} F_B[(K_1\otimes I_A)(|\psi_{SA}\rangle\langle \psi_{SA}|), (K_2\otimes I_A)(|\psi_{SA}\rangle\langle \psi_{SA}|)]\\
=&\min_{|\psi_{SA}\rangle}\max_{U_{ES2}}F_B[U_{ES1}\otimes I_A(|0_E\rangle\langle0_E|\otimes |\psi_{SA}\rangle\langle \psi_{SA}|) U^\dagger_{ES1}\otimes I_A,U_{ES2}\otimes I_A(|0_E\rangle\langle0_E|\otimes |\psi_{SA}\rangle\langle \psi_{SA}|) U^\dagger_{ES2}\otimes I_A]\\
=&\min_{|\psi_{SA}\rangle}\max_{U_{ES2}}F_B[|0_E\rangle\langle0_E|\otimes |\psi_{SA}\rangle\langle \psi_{SA}|, U_{ES1}^\dagger U_{ES2}\otimes I_A(|0_E\rangle\langle0_E|\otimes |\psi_{SA}\rangle\langle \psi_{SA}|) U^\dagger_{ES2}U_{ES1}\otimes I_A]\\
=&\min_{|\psi_{SA}\rangle}\max_{U_{ES2}}|\langle\psi_{SA}|\langle0_E| U_{ES1}^\dagger U_{ES2}\otimes I_A |0_E\rangle|\psi_{SA}\rangle|\\
=&\min_{|\psi_{SA}\rangle}\max_{U_{ES2}}|\langle\psi_{SA}|\langle0_E|\begin{bmatrix}
K_{W_d} & * & * & \cdots & *
\\
* & * & * & \cdots & *
\\
\vdots & & & & \vdots
\\
* & * & * & \cdots & *
\end{bmatrix}\otimes I_A|0_E\rangle|\psi_{SA}\rangle|\\
=&\min_{|\psi_{SA}\rangle}\max_{W_d}|\langle\psi_{SA}|K_{W_d}\otimes I_A|\psi_{SA}\rangle|\\
=&\min_{\rho^S}\max_{W_d} |Tr(\rho^SK_{W_d})|\\
=&\min_{\rho^S}\max_{W_d}|Tr(\sum_{ij}w_{ij}\rho^S F_{1i}^\dagger F_{2j})|\\
=&\min_{\rho^S}\max_{W_d}|\sum_{ij}w_{ij}Tr(\rho^SF_{1i}^\dagger F_{2j})|\\
=&\min_{\rho^S}\max_{W_d}|Tr[W_d^TM(\rho^S)]|,
\endaligned
\end{eqnarray}
\end{widetext}
where $\rho^S=Tr_A(\rho^{SA})$ and $M(\rho^S)$ is a $d\times d$ matrix with its $ij$-entry equals to $Tr(\rho^S F_{1i}^\dagger F_{2j})$. As $W_d$ is the first $d\times d$ submatrix of $W_p\in U(p)$, $\|W_d\|\leq 1$, here $\|\dot\|$ denotes the operator norm which equals to the maximum singular value. Conversely any $W_d$ satisfies $\|W_d\| \leq 1$ can be imbedded as a submatrix of a unitary\cite{Choi}, thus
\begin{widetext}
\begin{eqnarray}
\label{eq:minmax}
\aligned
&\min_{|\psi_{SA}\rangle} F_B[(K_1\otimes I_A)(|\psi_{SA}\rangle\langle \psi_{SA}|), (K_2\otimes I_A)(|\psi_{SA}\rangle\langle \psi_{SA}|)]
=&\min_{\rho^S}\max_{\|W_d\|\leq 1}|Tr[W_d^TM(\rho^S)]|.
\endaligned
\end{eqnarray}
\end{widetext}
Let $M(\rho^S)=UDV$ be the singular value decomposition of $M(\rho^S)$, where $U,V\in U(d)$ and $D=diag(s_1,s_2,\cdots,s_d)$, then $\max_{\|W_d\|\leq 1}|Tr[W_d^TM(\rho^S)]|=\|M(\rho^S)\|_1=\sum_{i=1}^d s_i$, the equality is achieved with $W_d^T=V^\dagger U^\dagger$.
This gives an analytical formula to compute the fidelity with any given pure input state

\begin{eqnarray}
\nonumber
\aligned
&F_B[(K_1\otimes I_A)(|\psi_{SA}\rangle\langle \psi_{SA}|), (K_2\otimes I_A)(|\psi_{SA}\rangle\langle \psi_{SA}|))\\
=&\|M(\rho^S)\|_1.
\endaligned
\end{eqnarray}

If we substitute $K_1$, $K_2$ with $K_x$ and $K_{x+dx}$, we get $F_B(\rho_x,\rho_{x+dx})=\|M(\rho^S)\|_1$, where $\rho_x=K_x\otimes I_A(|\psi_{SA}\rangle\langle \psi_{SA}|)$ and $\rho_{x+dx}=K_{x+dx}\otimes I_A(|\psi_{SA}\rangle\langle \psi_{SA}|)$, then using the connection between the Bures distance and the quantum Fisher information
\begin{equation}
\label{eq:BJ1}
d^2_{Bures}(\rho_x,\rho_{x+dx})=\frac{1}{4}J(\rho_x)dx^2,
\end{equation}
we get the analytical formula for quantum Fisher information of the extended channel with any pure input state $|\psi_{SA}\rangle$, $J(K_x\otimes I_A(\rho^{SA}))=\lim_{dx\rightarrow 0}\frac{8(1-\|M(\rho^S)\|_1)}{dx^2}$, where $\rho^S=Tr_A(|\psi_{SA}\rangle\langle \psi_{SA}|)$. Following the same line of argument, it can be shown that without auxillary system the same formula holds
\begin{eqnarray}
\aligned
&F_B[K_x(|\psi_{S}\rangle\langle \psi_{S}|), K_{x+dx}(|\psi_{S}\rangle\langle \psi_{S}|)]\\
=&\|M(\rho^S)\|_1,
 \endaligned
\end{eqnarray}
and $J(K_x(\rho^{S}))=\lim_{dx\rightarrow 0}\frac{8(1-\|M(\rho^S)\|_1)}{dx^2}$, just in this case $\rho^S=|\psi_{S}\rangle\langle \psi_{S}|$ is not the reduced state, but the pure input state $|\psi_{S}\rangle\langle \psi_{S}|$.

Note that when $W_d^T=V^\dagger U^\dagger$, $Tr[W_d^TM(\rho^S)]=\|M(\rho^S)\|_1$ is a positive real number, thus
 \begin{eqnarray}
 \aligned
 &\max_{\|W_d\|\leq 1}|Tr[W_d^TM(\rho^S)]|\\
 &=\max_{\|W_d\|\leq 1}Re\{Tr[W_d^TM(\rho^S)]\}\\
 &=\max_{\|W_d\|\leq 1}\frac{1}{2}Tr\{W_d^TM(\rho^S)+[W_d^TM(\rho^S)]^\dagger\}\\
 &=\max_{\|W_d\|\leq 1}\frac{1}{2}Tr\{\rho^SK_{W_d}+[\rho^SK_{W_d}]^\dagger\}\\
&=\max_{\|W_d\|\leq 1}\frac{1}{2}Tr[\rho^SK_{W_d}+K_{W_d}^\dagger\rho^S]\\
&=\max_{\|W_d\|\leq 1}\frac{1}{2}Tr[\rho^S(K_{W_d}+K_{W_d}^\dagger)],\\
 \endaligned
 \end{eqnarray}
 then
$\min_{|\psi_{SA}\rangle} F_B[(K_1\otimes I_A)(|\psi_{SA}\rangle\langle \psi_{SA}|), (K_2\otimes I_A)(|\psi_{SA}\rangle\langle \psi_{SA}|)]
=\min_{\rho^S}\max_{\|W_d\|\leq 1}\frac{1}{2}Tr[\rho^S(K_{W_d}+K_{W_d}^\dagger)],$
where $\rho^S=Tr_A(|\psi_{SA}\rangle\langle \psi_{SA}|)$ and $K_{W_d}=\sum_{i=1}^d\sum_{j=1}^dw_{ij}F_{1i}^\dagger F_{2j}$ with $w_{ij}$ as the $ij$-th entry of $W_d$. For the extended channel since all the reduced states $\rho^S$ forms a convex set, $\{W_d|\|W_d\|\leq 1\}$ is convex and compact and $\frac{1}{2}Tr[\rho^S(K_{W_d}+K^\dagger_{W_d})]$ is a bilinear function of entries of $\rho^S$ and $W_d$, thus from Sion's theorem\cite{Sion}, the sequence of min-max can be exchanged, so
\begin{eqnarray}
\label{eq:suppmin}
\aligned
&\min_{\rho^S}\max_{\|W_d\|\leq 1 } \frac{1}{2}Tr[\rho^S(K_{W_d}+K^\dagger_{W_d})]\\
&=\max_{\|W\|\leq 1 }\min_{\rho^S}\frac{1}{2}Tr[\rho^S(K_{W}+K^\dagger_{W})]\\
&=\max_{\|W\|\leq 1 }\frac{1}{2}\lambda_{\min}(K_{W}+K^\dagger_{W}),
\endaligned
\end{eqnarray}
here $\lambda_{\min}(K_{W}+K^\dagger_{W})$ denotes the minimum eigenvalue of $K_{W}+K^\dagger_{W}$. We used a different symbol $W$ after the exchange of min-max to emphasis that although the optimal value is not affected by the exchange of min-max, the optimal point that achieves the optimal value can change. If we substitute $K_1$ with $K_x$ and $K_2$ with $K_{x+dx}$, then we get the formula to compute the Bures angle between the extended channels
\begin{eqnarray}
\aligned
\label{eq:suppTE}
&\cos B(K_x\otimes I_A, K_{x+dx}\otimes I_A)\\
= &\min_{\rho^{SA}} F_B[(K_1\otimes I_A)(\rho^{SA}), (K_2\otimes I_A)(\rho^{SA})]\\
=&\max_{\|W\|\leq 1 }\frac{1}{2}\lambda_{\min}(K_{W}+K^\dagger_{W}),
 \endaligned
 \end{eqnarray}
which then gives the maximal quantum Fisher information for the extended channel $K_x\otimes I_A$
\begin{eqnarray}
\aligned
\label{eq:maxJextension}
&\max_{\rho^{SA}} J(K_x\otimes I_A(\rho^{SA}))\\
= &\lim_{dx\rightarrow 0}8\frac{1-\cos B(K_x\otimes I_A, K_{x+dx}\otimes I_A)}{dx^2}\\
=&\lim_{dx\rightarrow 0}\frac{8(1-\max_{\|W\|\leq 1 }\frac{1}{2}\lambda_{\min}(K_{W}+K^\dagger_{W})}{dx^2},
\endaligned
\end{eqnarray}
where $K_{W}=\sum_{ij}w_{ij}F_i^\dagger(x) F_j(x+dx)$.

\subsection{Connection with previous studies}
\label{sec:con}
The above formula includes previous studies as a special case.
Previous results  as in \cite{Fujiwara2008,Escher2011} state that for an extended channel $K_x\otimes I_A$ the maximal quantum Fisher information is given by
 \begin{equation}
  \max J=4\min_{\{\hat{F}_j(x)\}}\|\sum_{j=1}^d \dot{\hat{F}}^\dagger_j(x)\dot{\hat{F}}_j(x) \|
  \end{equation}
where the minimization is over all smooth representations of equivalent Kraus operators of the channel $K_x$. Note that this can be equivalently written as
\begin{widetext}
    \begin{eqnarray}
    \label{eq:maxJpre}
    \aligned
    \max J&=4\min_{\{\hat{F}_j(x)\}}\|\sum_{j=1}^d \dot{\hat{F}}^\dagger_j(x)\dot{\hat{F}}_j(x) \|\\
&=4\min_{\{\hat{F}_j(x)\}}\|\sum_{j=1}^d \lim_{dx\rightarrow 0}\frac{(\hat{F}^\dagger_j(x+dx)-\hat{F}^\dagger_j(x))}{dx}\frac{ (\hat{F}_j(x+dx)-\hat{F}_j(x))}{dx}\|\\
&=4\min_{\{\hat{F}_j(x)\}}\|\frac{2I-\sum_{j=1}^d(\hat{F}^\dagger_j(x)\hat{F}_j(x+dx)+\hat{F}^\dagger_j(x+dx)\hat{F}_j(x))}{dx^2}\|\\
&=4\frac{2-\max_{\{\hat{F}_j(x)\}} \lambda_{\min}[\sum_{j=1}^d (\hat{F}^\dagger_j(x)\hat{F}_j(x+dx)+\hat{F}^\dagger_j(x+dx)\hat{F}_j(x))]}{dx^2},
  \endaligned
  \end{eqnarray}
  \end{widetext}
  where the optimization is over all smooth representations of equivalent Kraus operators. In previous studies the equivalent Kraus operators are represented by $\hat{F}_j(x)=\sum_{i=1}^d \omega_{ji}(x)F_i(x)$ and $\hat{F}_j(x+dx)=\sum_{i=1}^d \omega_{ji}(x+dx)F_i(x+dx)$, where $\omega_{ji}(x)$ is $ji$-entry of $W_E(x)\in U(d)$, and $W_E(x)$ is required to be smooth with respect to $x$.
  It is easy to see that in this case Eq.(\ref{eq:maxJpre}) is just a special case of Eq.(\ref{eq:maxJextension}) when $W$ is restricted to taking the form $W_E^\dagger(x)W_E(x+dx)$.

We provide an example showing that the optimal $W$ in Eq.(\ref{eq:maxJextension}) can be non-unitary.


Consider two 8-dimensional channels with two Kraus operators each.
The first channel $K_1(dx)$ has the following operators:
\begin{align}
F_{11}&=
\text{diag}(\alpha,\alpha,\alpha,\alpha,\alpha,\alpha,\alpha,\alpha)
\\
F_{12}&=
\text{diag}(\beta,-\beta,\beta,-\beta,i\beta,-i\beta,i\beta,-i\beta)
\end{align}
The second channel $K_2(dx)$ has the following:
\begin{align}
F_{21}&=F_{11}
\\
F_{22}&=
\text{diag}(\beta,\beta,-\beta,-\beta,\beta,\beta,-\beta,-\beta)
\end{align}
where $\alpha=\sqrt{1-dx^2}$ and $\beta=|dx|$ are real.
Note that these two channels can be regarded as two neighbouring channels parameterized continuously in $x$, $\phi(x)=\left\{\begin{array}{cc}
      K_1(x)  & x\leq 0  \\
      K_2(x) & x>0 \\
          \end{array}\right..$

We let $W=\begin{bmatrix} w_{11} & w_{12} \\ w_{21} & w_{22}\end{bmatrix}$,
and compute $\frac{1}{2}\lambda_\text{min}(K_W+K_W^\dag)$, where
$K_W=w_{11} F_{11}^\dag F_{21} + w_{12}F_{11}^\dag F_{22} + w_{21}F_{12}^\dag F_{21} + w_{22}F_{12}^\dag F_{22}$.
First, note that if $W=\begin{bmatrix} 1&0\\0&0 \end{bmatrix}$, then $\frac{1}{2}\lambda_\text{min}(K_W+K_W^\dag)=\alpha^2$.

Now consider general $W$, since all matrices $F_{1i}^\dag F_{2j}$ are diagonal in this case, we will only need to look at the diagonal elements.
The diagonal elements of $K_W$ are
\begin{alignat}{4}
\alpha & (w_{11} \alpha & + & w_{12}\beta) &+ &\beta (w_{21} \alpha &+& w_{22} \beta)
\\
 &  &+ &&-& &+
\\
 &  &- &&+& &-
\\
 &  &- &&-& &-
\\
 &  &+ &&-i& &+
\\
 &  &+ &&+i& &+
\\
 &  &- &&-i& &-
\\
 &  &- &&+i& &-
\end{alignat}
Thus, $Tr(K_W)=8 \alpha^2 w_{11}$, thus $Tr(\frac{K_W+K_W^\dagger}{2})=8\alpha^2 Re(w_{11})$.
If $Re(w_{11})<1$, the average eigenvalue of $\frac{K_W+K_W^\dagger}{2}$ will be smaller than $\alpha^2$ thus the minimum eigenvalue will also be smaller than $\alpha^2$, which is worse than the case of $W=\begin{bmatrix} 1&0\\0&0 \end{bmatrix}$, thus not optimal.
So for optimal $W$, $w_{11}$ must be 1. 
With $w_{11}=1$, the most general $W$ with $\|W\|\leq 1$ is $W=\begin{bmatrix} 1&0\\0& re^{i \theta} \end{bmatrix}$ with $0\leq r\leq 1$. The diagonal elements of $K_W$ are then
$\alpha^2 \pm re^{i \theta} \beta^2$,
$\alpha^2 \pm i re^{i \theta} \beta^2$.
And the diagonal elements of $\frac{K_W+K_W^\dagger}{2}$ are
$$
\alpha^2 \pm \beta^2 r\cos(\theta),
\alpha^2 \pm \beta^2 r\sin(\theta).
$$
If $r>0$, one of them must be less than $\alpha^2$. So the optimal choice of $W$ in this case is $W=\begin{bmatrix} 1&0\\0&0 \end{bmatrix}$, which is not unitary.
And it is easy to check that the minimum value $\alpha^2$ for $F_B[(K_1(dx)\otimes I_A)(\rho^{SA}), (K_2(dx)\otimes I_A)(\rho^{SA})]$ in this case can be achieved by preparing the input state as the maximally entangled state $\frac{1}{\sqrt{8}}\sum_{i=1}^8|ii\rangle$.

We also note that as the set $\{W|\|W\|\leq 1\}$ is a convex set, this allowed a direct formulation with semi-definite programming as stated in the main text. While in previous studies $W$ is restricted to be unitary which does not form a convex set. To circumvent the difficulty previous study has to resort to the Lie algebra of the unitaries and formulated the semi-definite programming there instead\cite{Rafal2012}. That, however, comes with a cost on the computational complexity, which can be seen by comparing the size of the constraining matrices in the semi-definite programming: the constraining matrices in the semi-definite programming of Eq.(\ref{eq:sdp}) have the total size of $2d+m_1$, while previous study needs a size of $m_1+dm_2$\cite{Rafal2012}. The difference can be significant when the system gets large(note that for generic channels $d$ is in the order of $m_1m_2$). For example for $N$-qubit system, $m_1=m_2=2^N$, the difference quickly becomes large with the increase of $N$.

\subsection{Quantum Fisher information and optimal input states with dephasing noise and spontaneous emission}
We give detailed calculation for the maximum quantum Fisher information and optimal input states for two examples, one with dephasing noise and the other with spontaneous emission.

Consider a channel with dephasing noise $$K_x(\rho)=U(x)(\frac{1+\eta}{2}\rho+\frac{1-\eta}{2}\sigma_3\rho\sigma_3)U^\dagger(x),$$ where $U(x)=\exp(-i\frac{\sigma_3}{2}x), \eta\in [0,1]$.
  $\sigma_1=\left(\begin{array}{cc}
      0 & 1  \\
      1 & 0 \\
          \end{array}\right)$, $\sigma_2=\left(\begin{array}{cc}
      0 & -i  \\
      i & 0 \\
          \end{array}\right)$ and $\sigma_3=\left(\begin{array}{cc}
      1 & 0  \\
      0 & -1 \\
          \end{array}\right)$, $\eta\in [0,1]$.
In this case $F_1(x)=\sqrt{\frac{1+\eta}{2}}U(x)$, $F_2(x)=\sqrt{\frac{1-\eta}{2}}\sigma_3U(x)$. Let $\rho^{SA}$ be a pure input state for the extended channel $K_x\otimes I_A$, and $\rho^S=Tr_A(\rho^{SA})$, then
\begin{eqnarray}
\nonumber
\aligned
&M(\rho^S)\\
=&\left(\begin{array}{cc}
      Tr[\rho^SF_1^\dagger(x)F_1(x+dx)] & Tr[\rho^SF_1^\dagger(x)F_2(x+dx)]  \\
      Tr[\rho^SF_2^\dagger(x)F_1(x+dx)] & Tr[\rho^SF_2^\dagger(x)F_2(x+dx)] \\
          \end{array}\right)\\
&=\left(\begin{array}{cc}
      \frac{1+\eta}{2}C & \frac{\sqrt{1-\eta^2}}{2}D  \\
      \frac{\sqrt{1-\eta^2}}{2}D & \frac{1-\eta}{2}C \\
          \end{array}\right),
\endaligned
\end{eqnarray}
where $C=\rho^S_{11}\exp(-i\frac{dx}{2})+\rho^S_{22}\exp(i\frac{dx}{2}), D=\rho^S_{11}\exp(-i\frac{dx}{2})-\rho^S_{22}\exp(i\frac{dx}{2})$. Thus
$$M(\rho^S)M^\dagger(\rho^S)=\left(\begin{array}{cc}
      m_{11} & m_{12}  \\
      m_{21} &  m_{22}\\
          \end{array}\right),$$ where $m_{11}=\frac{(1+\eta)^2}{4}|C|^2+\frac{(1-\eta^2)}{4}|D|^2$, $m_{12}=\frac{\sqrt{1-\eta^2}}{4}[(1+\eta)CD^*+(1-\eta)C^*D]$, $m_{21}=\frac{\sqrt{1-\eta^2}}{4}[(1+\eta)C^*D+(1-\eta)CD^*]$, $m_{22}=\frac{(1-\eta)^2}{4}|C|^2+\frac{(1-\eta^2)}{4}|D|^2$.
          Denote $s_1$ and $s_2$ as singular values of $M(\rho^S)$, then $s^2_1+s^2_2=Tr(MM^\dagger)$ and $s_1s_2=|det(M)|$, thus
          \begin{eqnarray}
          \aligned
          &(s_1+s_2)^2\\
          =&Tr(MM^\dagger)+2|det(M)|\\
          =&\frac{1+\eta^2}{2}|C|^2+\frac{1-\eta^2}{2}|D|^2+\frac{1-\eta^2}{2}|C^2-D^2|\\
          =&(\rho^S_{11}+\rho^S_{22})^2-2\rho^S_{11}\rho^S_{22}\eta^2[1-\cos(dx)]\\
          =&1-4\rho^S_{11}\rho^S_{22}\eta^2\sin^2\frac{dx}{2}
\endaligned
\end{eqnarray}
To minimize $s_1+s_2$, we just need to maximize $\rho^S_{11}\rho^S_{22}$, which is achieved with $\rho^S_{11}=\rho^S_{22}=\frac{1}{2}$. Thus
$$\cos[B(K_x\otimes I_A,K_{x+dx}\otimes I_A)]=\sqrt{1-\eta^2\sin^2\frac{dx}{2}},$$ we then get the maximal quantum Fisher information $\lim_{dx\rightarrow 0}8\frac{1-\cos [B(K_x\otimes I_A,K_{x+dx}\otimes I_A)]}{dx^2}=\eta^2$, which is consistent with previous studies\cite{Jan2013,Rafal2014}. In this case the optimal initial state is any pure state $\rho^{SA}$ such that the reduced state $\rho^S$ satisfies $\rho^S_{11}=\rho^S_{22}=\frac{1}{2}$. The simplest choice is the pure state $\frac{|0\rangle+|1\rangle}{\sqrt{2}}$, i.e. in this case ancillary system is not necessary, which means that the Bures angle for $K_x$ and $K_{x+dx}$ remains the same
\begin{eqnarray}
\aligned
&\cos[B(K_x,K_{x+dx})]\\
=&\cos[B(K_x\otimes I_A,K_{x+dx}\otimes I_A)]\\
=&\sqrt{1-\eta^2\sin^2\frac{dx}{2}},
\endaligned
\end{eqnarray}
 thus the maximal quantum Fisher information under the evolution $K_x$ is also $\eta^2$. This is consistent with previous studies, however in our framework the optimal state comes out from the calculation naturally while in previous studies it depends on educated guess\cite{Rafal2014,Jan2013}.

For a channel with spontaneous emission, the Kraus operation takes form $K_x(\rho_0)=F_1(x)\rho_0 F^\dagger_1(x)+F_2(x)\rho_0 F_2^\dagger(x)$, where $F_1(x)=\left(\begin{array}{cc}
      1 & 0  \\
      0 & \sqrt{\eta} \\
          \end{array}\right)U(x),F_2(x)=\left(\begin{array}{cc}
      0 & \sqrt{1-\eta}  \\
      0 & 0 \\
          \end{array}\right)U(x),
U(x)=\exp(-i\frac{\sigma_3}{2}x)$. Again let $\rho^{SA}$ be a pure input state for the extended channel $K_x\otimes I_A$, and $\rho=Tr_A(\rho^{SA})$, then $$M(\rho)=\left(\begin{array}{cc}
      \rho_{11}e^{-i\frac{dx}{2}}+\eta \rho_{22}e^{i\frac{dx}{2}} & \sqrt{1-\eta}\rho_{21}e^{i(x+\frac{dx}{2})}  \\
      \sqrt{1-\eta}\rho_{12}e^{-i(x+\frac{dx}{2})} & (1-\eta)\rho_{22}e^{i\frac{dx}{2}} \\
          \end{array}\right).$$
          Denote $s_1$ and $s_2$ as the singular values of $M(\rho)$ we can similarly get
\begin{eqnarray}
\label{eqn-supplementary-spontaneous-1}
          \aligned
          &(s_1+s_2)^2\\     =&\rho^2_{11}+[\eta^2+(1-\eta)^2]\rho_{22}^2+2\eta\rho_{11}\rho_{22}\cos(dx)\\
          &+2(1-\eta)(|\rho_{21}|^2+|\rho_{11}\rho_{22}+\eta\rho_{22}^2e^{idx}-|\rho_{21}|^2|).
          \endaligned
\end{eqnarray}
First observe that
\begin{eqnarray}
\aligned
&|\rho_{21}|^2+|\rho_{11}\rho_{22}+\eta\rho_{22}^2e^{idx}-|\rho_{21}|^2|\\
&\geq ||\rho_{21}|^2+\rho_{11}\rho_{22}+\eta\rho_{22}^2e^{idx}-|\rho_{21}|^2|\\
&=|\rho_{11}\rho_{22}+\eta\rho_{22}^2e^{idx}|\\
&=\rho_{22}\sqrt{\rho_{11}^2+2\eta\rho_{11}\rho_{22}\cos dx+\eta^2\rho_{22}^2},
\endaligned
\end{eqnarray}
where the equality is achieved with $\rho_{21}=0$,
so \begin{eqnarray}
          \aligned
          &(s_1+s_2)^2\\
          \geq&\rho^2_{11}+[\eta^2+(1-\eta)^2]\rho_{22}^2+2\eta\rho_{11}\rho_{22}\cos(dx)
          \\&+2(1-\eta)\rho_{22}\sqrt{\rho_{11}^2+2\eta\rho_{11}\rho_{22}\cos dx+\eta^2\rho_{22}^2}\\
          =&[\sqrt{\rho_{11}^2+2\eta\rho_{11}\rho_{22}\cos dx+\eta^2\rho_{22}^2}+(1-\eta)\rho_{22}]^2\\
          \geq &[\sqrt{\frac{\eta+2\eta\sqrt{\eta}\cos dx+\eta^2}{(1+\sqrt{\eta})^2}}+\frac{1-\eta}{1+\sqrt{\eta}}]^2.
          \endaligned
\end{eqnarray}
The last inequality is achieved by substituting $\rho_{11}=1-\rho_{22}$ into the equation and find the minimum of a single variable function, which is saturated with $\rho_{11}=\frac{\sqrt{\eta}}{1+\sqrt{\eta}}$ and $\rho_{22}=\frac{1}{1+\sqrt{\eta}}$. The minimum is thus achieved at $\rho=\left(\begin{array}{cc}
      \frac{\sqrt{\eta}}{1+\sqrt{\eta}} & 0  \\
      0 & \frac{1}{1+\sqrt{\eta}} \\
          \end{array}\right)$, which gives the Bures angle
          \begin{eqnarray}
          \aligned
          &\cos[B(K_x\otimes I_A,K_{x+dx}\otimes I_A)]\\
          =&\sqrt{\frac{\eta+2\eta\sqrt{\eta}\cos dx+\eta^2}{(1+\sqrt{\eta})^2}}+\frac{1-\eta}{1+\sqrt{\eta}},
          \endaligned
          \end{eqnarray}

expanding both sides up to the second order of $dx$, we get the maximal quantum Fisher information $\lim_{dx\rightarrow 0}4\frac{B^2(K_x\otimes I_A,K_{x+dx}\otimes I_A)}{dx^2}=\frac{4\eta}{(1+\sqrt{\eta})^2}.$ The optimal initial state in this case is any pure state $\rho^{SA}$ with the reduced state equals to $\rho$, the simplest choice is $\sqrt{\frac{\sqrt{\eta}}{1+\sqrt{\eta}}}|00\rangle+\sqrt{\frac{1}{1+\sqrt{\eta}}}|11\rangle$.

The maximal quantum Fisher information for spontaneous emission without ancillary systems can also be calculated by imposing the state $\rho$ used in $M(\rho)$ to be pure:
$\rho=
\begin{bmatrix}
|a|^2 & a b^*\\
a^* b & |b|^2
\end{bmatrix}$ where $a,b$ are complex numbers and $|a|^2+|b|^2=1$.
Eq.~\eqref{eqn-supplementary-spontaneous-1} then becomes
\begin{eqnarray}
          \aligned
          (s_1+s_2)^2=&
|a|^4+|b|^4+2 \left[ \eta \cos dx + (1-\eta) \right] |a b|^2.
          \endaligned
\end{eqnarray}
Taking the derivative with respect to $|a|^2$ gives the extremal point $|a|^2=\frac{1}{2}$ corresponding to the minimum value
$$(s_1+s_2)^2=1+\frac{\eta}{2}(\cos dx -1)
=1-\frac{1}{4}\eta dx^2 +O(dx^4).
$$
(We know this is the minimum value by comparing to $|a|^2=0$.)
This gives the Bures angle
\begin{eqnarray}
\cos[B(K_x,K_{x+dx})]=
1-\frac{1}{8}\eta dx^2 +O(dx^4).
\end{eqnarray}
Finally,
we get the maximal quantum Fisher information without using ancilla to be $\lim_{dx\rightarrow 0}8\frac{1-\cos [B(K_x,K_{x+dx})]}{dx^2}=\eta$.
The optimal input state is any pure state with $\rho_{11}=\rho_{22}=\frac{1}{2}$, which can be achieved by $\frac{|0\rangle+\exp(i \theta)|1\rangle}{\sqrt{2}}$ for any $\theta \in \mathbb{R}$.


\end{document}